\expandafter\ifx\csname LaTeX\endcsname\relax
      \let\maybe\relax
\else \immediate\write0{}
      \message{You need to run TeX for this, not LaTeX}
      \immediate\write0{}
      \makeatletter\let\maybe\@@end
\fi
\maybe

\magnification=\magstephalf

\hsize=5.25truein
\vsize=8.3truein
\hoffset=0.37truein

\newdimen\frontindent \frontindent=.45truein
\newdimen\theparindent \theparindent=20pt


\let\em=\it

\font\tencsc=cmcsc10
\font\twelvebf=cmbx10 scaled 1200
\font\bmit=cmmib10  \font\twelvebmit=cmmib10 scaled 1200
\font\sixrm=cmr6 \font\sixi=cmmi6 \font\sixit=cmti8 at 6pt

\font\eightrm=cmr8  \let\smallrm=\eightrm
\font\eighti=cmmi8  \let\smalli=\eighti
\skewchar\eighti='177
\font\eightsy=cmsy8
\skewchar\eightsy='60
\font\eightit=cmti8
\font\eightsl=cmsl8
\font\eightbf=cmbx8
\font\eighttt=cmtt8
\def\eightpoint{\textfont0=\eightrm \scriptfont0=\fiverm 
                \def\rm{\fam0\eightrm}\relax
                \textfont1=\eighti \scriptfont1=\fivei 
                \def\mit{\fam1}\def\oldstyle{\fam1\eighti}\relax
                \textfont2=\eightsy \scriptfont2=\fivesy 
                \def\cal{\fam2}\relax
                \textfont3=\tenex \scriptfont3=\tenex 
                \def\it{\fam\itfam\eightit}\let\em=\it
                \textfont\itfam=\eightit
                \def\sl{\fam\slfam\eightsl}\relax
                \textfont\slfam=\eightsl
                \def\bf{\fam\bffam\eightbf}\relax
                \textfont\bffam=\eightbf \scriptfont\bffam=\fivebf
                \def\tt{\fam\ttfam\eighttt}\relax
                \textfont\ttfam=\eighttt
                \setbox\strutbox=\hbox{\vrule
                     height7pt depth2pt width0pt}\baselineskip=9pt
                \let\smallrm=\sixrm \let\smalli=\sixi
                \rm}


\catcode`@=11
 
\def\vfootnote#1{\insert\footins\bgroup\eightpoint
     \interlinepenalty=\interfootnotelinepenalty
     \splittopskip=\ht\strutbox \splitmaxdepth=\dp\strutbox
     \floatingpenalty=20000
     \leftskip=0pt \rightskip=0pt \parskip=1pt \spaceskip=0pt \xspaceskip=0pt
     \everydisplay={}
     \smallskip\textindent{#1}\footstrut\futurelet\next\fo@t}
 
\newcount\notenum

\def\note{\global\advance\notenum by 1
    \edef\n@tenum{$^{\the\notenum}$}\let\@sf=\empty
    \ifhmode\edef\@sf{\spacefactor=\the\spacefactor}\/\fi
    \n@tenum\@sf\vfootnote{\n@tenum}}


\tabskip1em

\newtoks\pream \pream={#\strut}
\newtoks\lpream \lpream={&#\hfil}
\newtoks\rpream \rpream={&\hfil#}
\newtoks\cpream \cpream={&\hfil#\hfil}
\newtoks\mpream \mpream={&&\hfil#\hfil}

\newcount\ncol \def\ncolp{\advance\ncol by 1}
\def\atalias#1{
    \ifx#1l\edef\xpream{\pream={\the\pream\the\lpream}}\xpream\ncolp\fi
    \ifx#1r\edef\xpream{\pream={\the\pream\the\rpream}}\xpream\ncolp\fi
    \ifx#1c\edef\xpream{\pream={\the\pream\the\cpream}}\xpream\ncolp\fi}
\catcode`@=\active

\def\taborl#1{\omit\unskip#1\hfil}
\def\taborc#1{\omit\hfil#1\hfil}
\def\taborr#1{\omit\hfil#1}
\def\multicol#1{\multispan#1\let\omit\relax}

\def\table#1\par{\midinsert\offinterlineskip\everydisplay{}
    \let@\atalias \let\l\taborl \let\r\taborr \let\c\taborc
    \def\space{\noalign{\vskip2pt}}
    \def\tablerule{\omit&\multispan{\the\ncol}\hrulefill\cr}
    \def\onerule{\space\space\tablerule\space\space}
    \def\tworules{\space\space\tablerule\space\tablerule\space\space}
    \def\annot##1\\{&\multispan{\the\ncol}##1\hfil\cr}
    \def\\{\let\\=\cr
           \edef\xpream{\pream={\the\pream\the\mpream}}\xpream
           \edef\starthalign{$$\vbox\bgroup\halign\bgroup\the\pream\cr}
           \starthalign
           \annot\hfil\tencsc Table #1\\ \noalign{\medskip}}
    \let\par\endtable}

\edef\endtable{\noalign{\vskip-\bigskipamount}\egroup\egroup$$\endinsert}

\let\plainmidinsert=\midinsert
\def\eightpttable{\def\midinsert{\let\midinsert=\plainmidinsert
    \plainmidinsert\eightpoint\tabskip 1em}\table}



\newif\iftitlepage

\def\raggedright{\rightskip 0pt plus .2\hsize\relax}

\let\caret=^ \catcode`\^=13 \def^#1{\ifmmode\caret{#1}\else$\caret{#1}$\fi}

\def\title#1\par{\vfill\supereject\begingroup
                 \global\titlepagetrue
                 \leftskip=\frontindent\parindent=0pt\parskip=0pt
                 \frenchspacing \eqnum=0
                 \gdef\runningtitle{#1}
                 \null\vskip-22.5pt\copy\volbox\vskip18pt
                 {\titlestyle#1\bigskip}}
\def\titlestyle{\raggedright\bf\twelvebf\textfont1=\twelvebmit
                \let\smallrm=\tenbf \let\smalli=\bmit
                \baselineskip=1.2\baselineskip}
\def\shorttitle#1\par{\gdef\runningtitle{#1}}
\def\author#1\par{{\raggedright#1\medskip}}

\def\shortauthor#1\par{\gdef\runningauthors{#1}}

\def\affil#1\par{{\raggedright\it#1\smallskip}}
\def@#1{\ifhmode\qquad\fi\leavevmode\llap{^{#1}}\ignorespaces}
\def\abstract{\smallskip\medskip{\bf Abstract: }}

\def\maybebreak#1{\vskip0pt plus #1\hsize \penalty-500
                  \vskip0pt plus -#1\hsize}

\def\maintextmode{\leftskip=0pt\parindent=\theparindent
                  \parskip=\smallskipamount\nonfrenchspacing}

\def\maintext#1\par{\bigskip\medskip\maintextmode\noindent}

\newcount\secnum
\def\section#1\par{\ifnum\secnum=0\medskip\maintextmode\fi
    \advance\secnum by 1 \bigskip\maybebreak{.1}
    \subsecnum=0
    \hang\noindent\hbox to \parindent{\bf\the\secnum.\hfil}{\bf#1}
    \smallskip\noindent}

\newcount\subsecnum
\def\subsection#1\par{\ifnum\subsecnum>0\medskip\maybebreak{.1}\fi
    \advance\subsecnum by 1
    \hang\noindent\hbox to \parindent
       {\it\the\secnum.\the\subsecnum\hfil}{\it#1}
    \par\noindent}

\def\references\par{\bigskip\maybebreak{.1}\parindent=0pt
    \everypar{\hangindent\theparindent\hangafter1}
    \leftline{\bf References}\smallskip}

\def\appendix#1\par{\bigskip\maybebreak{.1}\maintextmode
    \advance\secnum by 1 \bigskip\maybebreak{.1}
    \leftline{\bf Appendix: #1}\smallskip\noindent}

\def\bye{\endgroup\vfill\supereject\end}


\newbox\volbox
\setbox\volbox=\vbox{\hsize=.5\hsize \raggedright
       \sixit\baselineskip=7.2pt \noindent
       The Nature of Elliptical Galaxies,
       Proceedings of the Second Stromlo Symposium,
       Eds.\ M.~Arnaboldi, G.S.~Da~Costa \& P.~Saha}


\input epsf

\def\figureps[#1,#2]#3.{\midinsert\parindent=0pt\eightpoint
    \vbox{\epsfxsize=#2\centerline{\epsfbox{#1}}}
    \def\par{\endgraf\endinsert}{\bf Figure#3.}}

\def\figuretwops[#1,#2,#3]#4.{\midinsert\parindent=0pt\eightpoint
    \vbox{\centerline{\epsfxsize=#3\epsfbox{#1}\hfil
                      \epsfxsize=#3\epsfbox{#2}}}
     \def\par{\endgraf\endinsert}{\bf Figure#4.}}

\def\figurespace[#1]#2.{\midinsert\parindent=0pt\eightpoint
    \vbox to #1 {\vfil\centerline{\tenit Stick Figure#2 here!}\vfil}
    \def\par{\endgraf\endinsert}{\bf Figure#2.}}


\headline={\iftitlepage\hfil\else
              \ifodd\pageno\hfil\tensl\runningtitle
                    \kern1pc\tenbf\folio
               \else\tenbf\folio\kern1pc
                    \tensl\runningauthors\hfil\fi
           \fi}
\footline{\iftitlepage\tenbf\hfil\folio\hfil\else\hfil\fi}
\output={\plainoutput\global\titlepagefalse}


\newcount\eqnum
\everydisplay{\puteqnum}  
\def\puteqnum#1$${#1\global\advance\eqnum by 1\eqno(\the\eqnum)$$}
\def\namethiseqn#1{\xdef#1{\the\eqnum}}

 
\newcount\mpageno
\mpageno=\pageno  \advance\mpageno by 1000
 
\def\advancepageno{\global\advance\pageno by 1
                   \global\advance\mpageno by 1 }


\def\LaTeX{{\rm L\kern-.36em\raise.3ex\hbox{\tencsc a}\kern-.15em
    T\kern-.1667em\lower.7ex\hbox{E}\kern-.125emX}}

\def\[#1]{\raise.2ex\hbox{[}#1\raise.2ex\hbox{]}}

\def\witchbox#1#2#3{\hbox{$\mathchar"#1#2#3$}}
\def\leqsim{\mathrel{\rlap{\lower3pt\witchbox218}\raise2pt\witchbox13C}}
\def\geqsim{\mathrel{\rlap{\lower3pt\witchbox218}\raise2pt\witchbox13E}}

\def\<#1>{\langle#1\rangle}


{\obeyspaces\gdef {\ }}

\catcode`@=12 \let\@=@ \catcode`@=13
\def\+{\catcode`\\=12\catcode`\$=12\catcode`\&=12\catcode`\#=12%
       \catcode`\^=12\catcode`\_=12\catcode`\~=12\catcode`\%=12%
       \catcode`\@=0\tt}
\def\({\endgraf\bgroup\let\par=\endgraf\parskip=0pt\vskip3pt
       \eightpoint \def\/{{\eightpoint$\langle$Blank line$\rangle$}}
       \catcode`\{=12\catcode`\}=12\+\obeylines\obeyspaces}
\def\){\vskip1pt\egroup\vskip-\parskip\noindent\ignorespaces}


\title The Enrichment of the Intracluster Medium

\shorttitle The Enrichment of the ICM

\author Brad K. Gibson^1 \& Francesca Matteucci^2

\affil @1 Mount Stromlo \& Siding Spring Observatories, Australia

\affil @2 Scuola Internazionale Superiore di Studi Avanzati, Italy

\shortauthor Gibson \& Matteucci

\section The Galaxy Models

Gibson's (1996) coupled photo-chemical evolution package was used to
construct a grid of elliptical galaxy models consistent with the present-day
colour-luminosity-metallicity relations.  The enrichment of the ICM, through
supernova (SN)-driven winds, was considered under two scenarios governing the
binding energy (BE)-mass-radius 
relations -- (i) the \it standard model\rm, which
simply uses the conventional present-day relation due to Saito (1979), and (ii)
the \it reduced BE model\rm, which presumes that earlier
winds may be established while the local BE is substantially lower 
(a factor of $\sim
5$ was chosen \it a posteriori\rm) than Saito's
relation would predict (e.g. perhaps due to trace star formation driving a wind
during a proto-galaxy's pre-collapse phase?).  This is almost certainly an
unphysical model, but our goal is simply to explore mechanisms, from within
this framework, for maximizing the gas mass ejected per galaxy to the ICM, in
order to determine which of the parameters must be stretched to their breaking
points before we can recover 100\% of the ICM gas through the
``process-and-ejected'' hypothesis of Trentham (1994).
Table 1 shows our galaxy
grid; six different initial gas masses (embedded within dark halos) are
covered.  Star formation is taken to be $\psi(t)=\nu {\rm M_g}(t)$.  Galactic
wind onset occurs at $t_{\rm GW}$.
The ratio of ejected gas mass to the present-day V-band luminosity is denoted
$m_{\rm g}^{\rm ej}/L_{\rm V}$.  Under the reduced BE model, $\nu$ must be
decreased dramatically in order to delay $t_{\rm GW}$ until present-day
photo-chemical properties are recovered.  The smaller BE 
means means far fewer SNe are needed to drive the wind, which means, in
general, a much lower final L$_{\rm V}$, and hence the increased ratio in the
final column.

\table 1. Template Galaxy Models

@c @r @r @r @r @r @r \\
\tworules
&\c{M$_{\rm g}(0)$} & \c{$\nu$} & \c{$t_{\rm GW}$} & \c{$m_{\rm g}^{\rm ej}/L_{\rm V}$} & \c{$\nu$} & \c{$t_{\rm GW}$} & \c{$m_{\rm g}^{\rm ej}/L_{\rm V}$} \\
& & \multicol3\c{\it Standard Model} & \multicol3\c{\it Reduced BE Model} \\
\onerule
& 1.0e4  & 100.0  & 0.003  & 20.6$\quad$ & 2.4 & 0.007 & 386.9$\;\;$ \\
& 1.0e6  & 100.0  & 0.005  & 12.5$\quad$ & 2.3 & 0.011 & 239.7$\;\;$ \\
& 5.0e7  & 126.1  & 0.006  &  9.3$\quad$ & 3.3 & 0.013 & 150.7$\;\;$ \\
& 1.0e9  &  82.8  & 0.013  &  8.1$\quad$ & 2.8 & 0.022 & 107.5$\;\;$ \\
& 5.0e10 &  50.1  & 0.034  &  4.7$\quad$ & 1.4 & 0.332 &  18.4$\;\;$ \\
& 1.0e12 &  42.2  & 0.060  &  3.2$\quad$ & 0.8 & 1.391 &   7.8$\;\;$ \\
\onerule

\section The Cluster Models

The galaxy models of Table 1 were used in integrating over a two component
luminosity function (LF), 
with faint-end slope $\alpha_1$ for M$_{\rm V}<-17.3$ and
$\alpha_2$ for M$_{\rm V}>-17.3$, in order to determine the gas mass
fraction of the ICM which originated from within cluster
ellipticals;  
the numerical technique is outlined in
Gibson \& Matteucci (1997).  Three different LFs are shown in
Table 2, for both the standard and reduced BE galaxy models.  
We have purposely considered the steepest dwarf LFs seen by De Propris
\it et al. \rm 
(1995), in an attempt to recover 100\% of the ICM gas
via the ``wind''  scenario.  A built-in constraint is Melnick \it et al.\rm's
(1977) observation that the 
maximum contribution to a cluster's diffuse light, from dwarfs 
with M$_{\rm
V}>-17$ is $\sim 25$\%.  Because of this
constraint, all models with $\alpha_2\leqsim -1.6$ had to have artificially
increased values of M$_{\rm V}^{\rm min}$,the lower luminosity limit to the
integration, which is reflected in Table 2.  The other lower limits in Table 1
are essentially arbitrarily chosen.
The standard model predicts that
$\sim 20$\% of the ICM gas originates from ellipticals under the 
galactic wind 
scenario; the remaining $\sim 80$\% is primordial, in agreement with
Matteucci \& Vettolani (1988).  A wide range of parameter space was covered,
but we found that the only way in which to account for $\sim 100$\% of the ICM
gas was to adopt the unphysical reduced BE model, with a BE-mass-radius
relation $\sim 5\times$ lower than the canonical present-day relation predicts.

\table 2. Cluster Models

@r @r @r @r @r @r @r @r \\
\tworules
&\c{$\alpha_1$} & \c{$\alpha_2$} & \c{M$_{\rm V}^{\rm min}$} & \c{f$_{\rm g}^{\rm ICM}$} & \c{f$_{\rm g}^{\rm M>-17}$} & \c{M$_{\rm V}^{\rm min}$} & \c{f$_{\rm g}^{\rm ICM}$} & \c{f$_{\rm g}^{\rm M>-17}$} & \\
& & & \multicol3\c{\it Standard Model} & \multicol3\c{\it Reduced BE Model} \\
\onerule
& -1.45 & -1.45$\;$ &  -1.8$\quad$ & 0.20 & 0.25$\;\;$ &   1.3$\;$ & 0.83 & 0.43$\;\;$ \\
& -1.45 & -1.90$\;$ & -13.6$\quad$ & 0.20 & 0.35$\;\;$ & -13.8$\;$ & 0.86 & 0.50$\;\;$ \\
& -1.45 & -2.20$\;$ & -15.0$\quad$ & 0.20 & 0.32$\;\;$ & -14.9$\;$ & 0.81 & 0.46$\;\;$ \\
\onerule

\references

De Propris, R. \it et al. \rm 1995, ApJ, 450, 534

Gibson, B.K. 1996, ApJ, 468, 167

Gibson, B.K. \& Matteucci, F. 1997, 473, in press

Matteucci, F. \& Vettolani, G. 1988, A\&A, 202, 21

Melnick, J., White, S.D.M. \& Hoessel, J. 1977, MNRAS, 180, 207

Saito, M. 1979, PASJ, 31, 181

Trentham, N. 1994, Nature, 372, 157

\bye